\documentclass[journal]{IEEEtran}

\usepackage[T1]{fontenc}
\usepackage{cite}
\usepackage{amsmath,amssymb}
\usepackage{graphicx}
\usepackage{booktabs}
\usepackage{array}
\usepackage{url}
\usepackage{microtype}
\usepackage[hidelinks]{hyperref}
\usepackage{balance}

\graphicspath{{figures/}}
\newcommand{\reportfigure}[4]{%
  \begin{figure}[!t]
    \centering
    \IfFileExists{figures/#1}{%
      \includegraphics[width=#4,keepaspectratio]{figures/#1}%
    }{%
      \fbox{\parbox[c][1.05in][c]{0.88\columnwidth}{%
        \centering\small\textbf{Figure placeholder}\\[3pt]
        \texttt{\detokenize{#1}}\\[5pt]
        Crop the corresponding figure from the project report and save it
        using this filename in the \texttt{figures/} folder.%
      }}%
    }%
    \caption{#2}
    \label{#3}
  \end{figure}%
}

\begin{document}

\title{A Proof-of-Concept Study of Wi-Fi Channel State Information-Based Fall Detection Using ESP32 Devices}

\author{Amritha S.~S., Anhajul S.~Rahman, Mohammed Bin~Saleem, 
Vaishnav~Sankar, and Jerrin Thomas Panachakel%
\thanks{All authors are with the Department of Electronics and 
Communication Engineering, College of Engineering Trivandrum, 
Thiruvananthapuram 695016, Kerala, India 
(e-mail: jerrin.panachakel@cet.ac.in). }%
}

\markboth{Wi-Fi CSI-Based Fall Detection Using ESP32 Devices}{}
\maketitle

\begin{abstract}
This paper reports a small-scale proof-of-concept study of contactless fall detection using Wi-Fi channel state information (CSI) collected with two ESP32 devices. One device acts as the transmitter and the other as the receiver, with a separation of 3.5~m. The evaluated signal-processing pipeline uses CSI amplitude, second-level discrete wavelet transform (DWT) denoising with a Symlet-10 wavelet, z-score normalization, and fixed-length segmentation. Wavelet-scattering coefficients are used as features, and MATLAB's automated classifier-selection workflow identifies a LogitBoost ensemble with a minimum leaf size of four and a learning rate of 0.022052. The proposed classifier achieved an accuracy of 93.10\%, with a precision of 100\%, recall of 87.50\%, and F1-score of 93.33\%. These results demonstrate the potential of Wi-Fi CSI-based sensing for contactless fall detection using low-cost ESP32 hardware.
\end{abstract}


\section{Introduction}
Falls are an important source of injury and loss of independence among older adults. The consequences can be aggravated when a person cannot get up or summon assistance after falling. Automatic detection may help reduce the time between a fall and the arrival of assistance, although a detection system must operate reliably in ordinary living environments to be useful \cite{rubenstein2006falls}.

Conventional approaches commonly use wearable sensors or cameras. Wearable systems require the user to carry or wear a device and keep it charged, while camera-based systems raise privacy concerns and depend on suitable camera placement and lighting. Device-free Wi-Fi sensing offers a different approach: changes in the wireless propagation channel caused by movement can be inferred from channel state information (CSI), without requiring the monitored person to wear a sensor or be recorded by a camera \cite{ma2019wifi,wu2017devicefree}.

CSI provides a more fine-grained description of a wireless channel than a received signal-strength indicator because it represents the channel response at individual OFDM subcarriers. Human movement changes the propagation paths and therefore changes the measured channel response over time. Previous studies have used commodity Wi-Fi equipment for unobtrusive fall detection, including RT-Fall, WiFall, and FallDeFi \cite{wang2017rtfall,wang2017wifall,palipana2017falldefi}. More recent work has explored deep-learning approaches and broader multi-environment data collection for CSI-based fall detection \cite{chu2023deeplearning}.

The present work examines whether an inexpensive pair of ESP32 devices can provide CSI measurements suitable for distinguishing simulated falls from walking. A custom Python collection utility alongside Espressif's CSI functionality, followed by offline signal processing and machine-learning analysis were performed. ESP32-based CSI acquisition is supported by the device's Wi-Fi driver and CSI interface, which expose complex subcarrier channel responses to the receiving application \cite{espressifcsi,atif2020wiesp}.

The principal contributions of this paper: (1) a low-cost, two-node ESP32 CSI acquisition arrangement; (2) an amplitude preprocessing pipeline using wavelet denoising and z-score normalization; and (3) a preliminary fall-versus-walk classification experiment using wavelet-scattering features. 

\section{Related Work}
\subsection{Wi-Fi CSI for Device-Free Sensing}
Ma \emph{et al.} provide a broad survey of CSI-based Wi-Fi sensing and describe its use in detection, recognition, and estimation tasks \cite{ma2019wifi}. CSI-based methods exploit the fact that wireless propagation is affected by reflection, scattering, and multipath. A time sequence of channel responses can therefore encode changes in the surrounding environment, including changes caused by human movement. Wu \emph{et al.} distinguish pattern-based techniques, which learn directly from signal variations, from model-based techniques that incorporate radio-propagation models \cite{wu2017devicefree}.

\subsection{Wi-Fi-Based Fall Detection}
RT-Fall uses commodity Wi-Fi CSI and emphasizes automatic segmentation and low-latency detection. Its method makes use of phase differences across antennas as well as amplitude information \cite{wang2017rtfall}. WiFall also uses CSI for device-free fall detection and evaluates one-class support-vector-machine and random-forest approaches in indoor scenarios \cite{wang2017wifall}. FallDeFi uses time-frequency information derived from CSI and feature selection to improve fall detection when environmental conditions change \cite{palipana2017falldefi}.

Chu \emph{et al.} evaluate Wi-Fi CSI collection tools and propose a deep-learning-based fall detector using a dataset containing more than 700 CSI samples from four indoor environments. Their study reports accuracy above 96\% across the combined environments and higher accuracy for selected environment combinations \cite{chu2023deeplearning}. This broader data-collection context is relevant to the present study: CSI-based systems can be sensitive to room layout and other environmental conditions, so results obtained in a single controlled room should not be interpreted as evidence of environment-independent performance.

\subsection{ESP32-Based CSI Acquisition}
Atif \emph{et al.} describe Wi-ESP, an ESP32-based tool for collecting CSI for device-free Wi-Fi sensing applications \cite{atif2020wiesp}. Espressif's ESP-IDF documentation describes CSI as channel-frequency-response information estimated for received packets and represented using real and imaginary components for subcarriers \cite{espressifcsi}. These capabilities provide the acquisition basis for the low-cost sensing arrangement considered here.

\section{System and Methodology}
\subsection{System Configuration}
The acquisition arrangement uses two ESP32 microcontroller boards. One board operates as a transmitter and the other as a receiver. The devices were placed approximately 3.5~m apart, with the activity performed in the region between them. 

The ESP32 receiver obtains CSI associated with received Wi-Fi packets. For a subcarrier $k$ and time index $t$, the complex channel response can be represented as
\begin{equation}
H_k(t)=H_{k,\mathrm{R}}(t)+jH_{k,\mathrm{I}}(t),
\label{eq:csi-complex}
\end{equation}
where $H_{k,\mathrm{R}}(t)$ and $H_{k,\mathrm{I}}(t)$ denote the real and imaginary components. The amplitude used in the described classification pipeline is
\begin{equation}
A_k(t)=|H_k(t)|=\sqrt{H_{k,\mathrm{R}}^2(t)+H_{k,\mathrm{I}}^2(t)}.
\label{eq:csi-amplitude}
\end{equation}

\subsection{CSI Collection}
A custom Python utility was used together with Espressif's CSI collection functionality. The receiver recorded raw CSI-related packet information, from which the channel response components and amplitude traces were extracted.

\subsection{Experimental Dataset}
Data were collected in a spacious indoor room with relatively few furnishings. The activity window was five-seconds and a single participant performed each fall or walking sample; empty-room samples were recorded without a person in the monitored area. Fall and walking movements were performed in different directions or styles.

\subsection{Signal Preprocessing}
The recorded CSI is represented by real and imaginary components. The amplitude for each subcarrier is obtained using (\ref{eq:csi-amplitude}). 

\subsubsection{Denoising}
Two denoising attempts are compared: the first is a fourth-order Butterworth low-pass filter with a stated cutoff frequency of 10~Hz.  The second approach is a second-level discrete wavelet transform using the Symlet-10 wavelet. 

\subsubsection{Normalization and Fixed-Length Selection}
After denoising, the amplitude records were normalized using z-scores. For a record with mean $\mu$ and standard deviation $\sigma$, the normalized amplitude is
\begin{equation}
Z_k(t)=\frac{A_k(t)-\mu_k}{\sigma_k},
\label{eq:zscore}
\end{equation}
where $\mu_k$ and $\sigma_k$ are the mean and standard deviation of the amplitude values for the record. 

\subsection{Feature Extraction and Classifier Selection}
Wavelet scattering applies cascaded wavelet transforms and produces time-frequency representations that can be supplied to a classifier \cite{mathworkswavelet}. 

MATLAB's \texttt{fitcauto} workflow was used for automated classification-model selection and hyperparameter optimization \cite{mathworksfitcauto}. The best-performing candidate was an ensemble classifier using LogitBoost, with a minimum leaf size of 4 and a learning rate of 0.022052. 

\subsection{Evaluation Measures}
The task is binary classification of fall versus walking. Taking fall as the positive class, the standard measures are
\begin{align}
\mathrm{Accuracy} &= \frac{TP+TN}{TP+TN+FP+FN},\\
\mathrm{Precision} &= \frac{TP}{TP+FP},\\
\mathrm{Recall} &= \frac{TP}{TP+FN},\\
F_1 &= \frac{2\,\mathrm{Precision}\,\mathrm{Recall}}
{\mathrm{Precision}+\mathrm{Recall}},
\end{align}
where $TP$, $TN$, $FP$, and $FN$ denote true positives, true negatives, false positives, and false negatives, respectively.

\section{Results}

\subsection{Classifier Selection}

The automated classifier-selection procedure implemented in MATLAB identified a LogitBoost ensemble as the best-performing model for the fall-detection task. The selected ensemble was configured with a minimum leaf size of 4 and a learning rate of 0.022052. The LogitBoost algorithm combines multiple weak learners through an iterative boosting process to construct a classification model.

\subsection{Fall-Detection Performance}

The classification performance was evaluated using the confusion matrix presented in Table~\ref{tab:confusion}. The evaluation comprised 29 instances, including 16 fall events and 13 walking events. Of the 16 fall events, 14 were correctly classified, while two were misclassified as walking. All 13 walking events were correctly classified, with no false-positive fall detections.

\begin{table}[!t]
\caption{Confusion Matrix for Fall Detection}
\label{tab:confusion}
\centering
\begin{tabular}{lrr}
\toprule
\textbf{Actual Class} & \textbf{Predicted Fall} & \textbf{Predicted Walk}\\
\midrule
Fall  & 14 & 2\\
Walk  & 0  & 13\\
\bottomrule
\end{tabular}
\end{table}

The overall classification accuracy was 93.10\%. The classifier achieved a precision of 100\%, indicating that all events predicted as falls corresponded to actual fall events in the evaluated set. The recall was 87.50\%, reflecting the identification of 14 out of 16 actual falls. The resulting F1-score was 93.33\%. A summary of the classification performance is provided in Table~\ref{tab:metrics}.

\begin{table}[!t]
\caption{Classification Performance of the Selected LogitBoost Model}
\label{tab:metrics}
\centering
\begin{tabular}{lc}
\toprule
\textbf{Performance Measure} & \textbf{Value}\\
\midrule
Accuracy  & 93.10\%\\
Precision & 100.00\%\\
Recall    & 87.50\%\\
F1-score  & 93.33\%\\
Evaluated instances & 29\\
\bottomrule
\end{tabular}
\end{table}

\section{Discussion}

The experimental results demonstrate the feasibility of using Wi-Fi channel state information for contactless fall detection with low-cost ESP32 hardware. The selected LogitBoost ensemble achieved an accuracy of 93.10\%, with a precision of 100\% and an F1-score of 93.33\% on the evaluated fall and walking instances. These results indicate that CSI amplitude variations, when appropriately processed and represented using wavelet-scattering features, can provide discriminative information for distinguishing falling events from walking activity.

An important observation is the difference between precision and recall. Although no walking event was incorrectly classified as a fall, two of the 16 actual fall events were missed by the classifier. The resulting recall of 87.50\% indicates that the model did not identify every fall event in the evaluation. This distinction is particularly important in safety-oriented applications, where missed falls may have serious consequences. Consequently, further improvements in fall sensitivity are necessary before the approach can be considered for practical monitoring applications.

The results should also be interpreted in the context of the experimental scale. The confusion matrix represents 29 evaluated instances involving two activity classes: falling and walking. The absence of false-positive detections among the 13 walking instances does not establish a consistently low false-alarm rate under more varied everyday activities. Furthermore, the dataset was collected in a controlled indoor environment and involved a relatively small number of young participants. The effect of differences in body characteristics, movement patterns, room configurations, furniture arrangements, and transmitter--receiver positioning therefore remains to be investigated.

\section{Conclusion}

This study presents a contactless fall-detection approach based on Wi-Fi channel state information collected using two ESP32 devices. The proposed processing pipeline combines CSI amplitude extraction, second-level Symlet-10 discrete wavelet transform denoising, z-score normalization, fixed-length segmentation, and wavelet-scattering feature extraction. A LogitBoost ensemble classifier, selected through MATLAB's automated classifier-selection workflow, was employed to distinguish fall events from walking activity.

The selected classifier achieved an accuracy of 93.10\%, a precision of 100\%, a recall of 87.50\%, and an F1-score of 93.33\% on the evaluated set of 29 instances. These findings demonstrate the potential of commodity Wi-Fi devices and CSI-based signal analysis as a low-cost, contactless approach to fall detection.

Although the results are promising, further investigation is required to establish the robustness and generalisability of the approach. In particular, larger datasets, more diverse activities, subject-independent testing, and evaluations across different indoor environments are necessary. Addressing these challenges will be important in advancing ESP32-based Wi-Fi sensing towards reliable fall-monitoring applications.

\balance
\bibliographystyle{IEEEtran}
\bibliography{references}
\end{document}